\documentclass[12pt]{article}
\usepackage{amssymb}
\usepackage{amstext}

\def\a{\alpha}
\def\ha{{\hat \alpha}}
\def\b{\beta}
\def\bM{{\bf M}}
{

\def\hf{{1 \over 2}}

\def\hr{{\hat \rho}}

\def\w{\wedge}

\def\be{\begin{equation}}
\def\ee{\end{equation}}
\def\bea{\begin{eqnarray}}
\def\eea{\end{eqnarray}}
\def\nn{\nonumber\\}

\begin{document}
\pagestyle{empty}
\begin{flushright}
\begin{tabular}{ll}
MCTP-04-68 & \\
ITFA-2004-61 & \\
hep-th/0412120 &
 \\ [.3in]
\end{tabular}
\end{flushright}
\begin{center}
{\Large {\bf{On topological F-theory}}} \\ [.5in]
{{Lilia Anguelova$\, {}^{1}$, Paul de Medeiros$\, {}^{1}$ and Annamaria Sinkovics$\, {}^{2}$
}} \\ [.3in]
$\,{}^{1}$\, {\emph{Michigan Center for Theoretical Physics, Randall Laboratory, \\
University of Michigan, Ann Arbor, MI 48109-1120, U.S.A.}} \\ [.3in]
$\,{}^{2}$\, {\emph{Institute for Theoretical Physics, University of Amsterdam \\ Valckenierstraat 65, 1018 XE Amsterdam, The Netherlands.}} \\ [.3in]
{\tt{anguelov@umich.edu}}, {\tt{pfdm@umich.edu}}, {\tt{sinkovic@science.uva.nl}}
\\ [.5in]

{\large{\bf{Abstract}}} \\ [.2in]
\end{center}
We consider the construction of a topological version of F-theory on a particular $Spin(7)$ 8-manifold which is
a Calabi-Yau 3-fold times a 2-torus. We write an action for this theory in eight dimensions and reduce it to lower dimensions using Hitchin's gradient flow method.  A symmetry of the eight-dimensional theory which follows from modular transformations of the torus induces duality transformations of the
variables of the topological A- and B-models.
We also consider target space form actions in the presence of background fluxes in six dimensions.
\clearpage
\pagestyle{plain}
\pagenumbering{arabic}

%%%%%%%%%%%%%%%%%%%%%%%%%%%%%%%%%%%%%%%%%%%%%%%%%%%%%%%%%%%%%%%%%%%%%%%%%%%%%%%%%%%%%%%%%%%%%%%%%%%%%%%%%%%%%%%%%%%%%%%%%%%%%%%%%%%%%%%%%%%%%%%%%%%%%%%%%%%

\section{\large{Introduction}}

Recently, a new theory called {\emph{topological M-theory}} was constructed on seven-manifolds with $G_2$ structure \cite{M} \cite{Z}. It is proposed as a unification of the A- and B-model topological string theories which are themselves related to counting maps from Riemann surfaces into Calabi-Yau three-folds. The topological M-theory is related to these topological string theories via dimensional reduction along the seventh dimension. This is similar to the relationship between 11-dimensional M-theory and the type II superstring theories in ten dimensions.
Evidence for the existence of topological M-theory was given also in \cite{GS, GV}.

Topological strings have many interesting applications. It has long been known that the topological
A- and B-models compute F-terms for compactifications of type IIB superstring theory on Calabi-Yau three-folds \cite{Fterms}. Recently it was also discovered that they are related to perturbative $N=4$ super Yang-Mills theory \cite{EW} and to the entropy of BPS black holes in four dimensions \cite{OSV}. This underscores the importance of understanding their non-perturbative formulation.

In this note we examine the possibility, already raised in \cite{M}, that a topological theory on eight-manifolds with $Spin(7)$ structure may also be constructed. The framework is analogous to 12-dimensional F-theory and so is named {\emph{topological F-theory}}. For simplicity, we will consider the eighth dimension to be compact (and in fact circular in the discussion of dualities).

The effective action for topological M-theory was obtained using Hitchin's formalism for volume functionals that are built out of stable forms \cite{Hitchin}. Within the cohomology class of a given stable closed 3- or 4-form in seven dimensions, the extrema of this action functional precisely correspond to Riemann metrics of $G_2$ holonomy.
We will adopt a similar strategy here to construct an action in eight dimensions.

We consider the eight-dimensional space to be a seven-manifold with $G_2$ structure fibred over a line interval or circle. We will find it convenient to think in terms of the Cayley 4-form that defines the $Spin(7)$ structure on this manifold.
Although this 4-form is not stable, it determines a $Spin(7)$ holonomy metric. We use the gradient flow constructed by Hitchin \cite{Hitchin} to relate this Cayley 4-form to the 3- and 4-forms associated with $G_2$ structure on the seven-manifold. We write an action functional in eight dimensions in terms of this special Cayley form which precisely reduces to the topological M-theory effective action \cite{M} on compactifying the eighth dimension.
By reducing on an additional circle, we rewrite the Cayley form in terms of the basic quantities of the topological A- and B-models, the K\"{a}hler form and complex structure.

It has recently been conjectured that there exists a duality, called {\emph{topological S-duality}}, exchanging A- and B-model topological
strings on the same Calabi-Yau three-fold target space \cite{NV}. It was related in \cite{S} to the S-duality of type IIB superstrings. Similarly to the connection of the latter with physical F-theory, it is natural to ask if our eight-dimensional construction
can be used to analyse topological S-duality.
A naive canonical
quantization of topological M-theory on the product space ${\bf{M}}_6 \times {\mathbb{R}}$
was performed in \cite{M}, and the canonically conjugate variables are eventually identified
with the real and imaginary parts of the 3-form $\Omega$ corresponding to the $SU(3)$ structure on ${\bf{M}}_6$. It was also suggested that S-duality
would exchange these two conjugate variables.
We therefore look for symmetries of the eight-dimensional theory which induce such an exchange in the reduced theory on ${\bf{M}}_6$.
We find that a change of variables, similar to the one involved in the S-duality between the topological string theories, is induced from modular transformations of the two-torus upon requiring invariance of the Cayley form that defines the $Spin(7)$ structure.
In this sense, the topological F-theory seems a real analogy to physical F-theory,
where the complex structure change of the extra torus induces S-duality in type IIB superstrings.

A point which remains to be resolved is the embedding of our topological S-duality
in physical string theory. In \cite{S} an embedding in superstring theory
was described, which allowed to deduce the appropriate dependence
on the string coupling constant. In Section 6 we include some discussion on introducing coupling constant dependence in our formalism, but a full treatment is left for future work.

To better understand the action of S-duality, it would be desirable to include background fluxes
in our construction. In this paper we make initial steps in this direction by considering action functionals involving fluxes, which are related to six-dimensional non-K\"{a}hler manifolds.
%%%%%%%%%%%%%%%%%%%%%%%%%%%%%%%%%%%%%%%%%%%%%%%%%%%%%%%%%%%%%%%%%%%%%%%%%%%%%%%%%%%%%%%%%%%%%%%%%%%%%%%%%%%%%%%%%%%%%%%%%%%%%%%%%%%%%%%%%%%%%%%%%%%%%%%%%%%

\section{\large{Eight-manifolds with Spin(7) holonomy}}

We begin by giving a short introduction to $Spin(7)$ structures on eight-dimensional manifolds. The first examples of metrics with $Spin(7)$ holonomy were given in \cite{BS} and a more general class of such metrics was given in \cite{Joyce}. In those references, together with \cite{Hitchins}, one can also find a more thorough introduction.

A $Spin(7)$ holonomy metric $g$ on an eight-dimensional manifold ${\bf{M}}_8$ is defined by the existence of a {\emph{Cayley}} 4-form $\Psi$ which is both closed
\be \label{dpsi}
d \Psi \; =\; 0 \; ,
\ee
and self-dual
\be
* \Psi \; =\; \Psi \; .
\ee
The expression for the metric $g$ in terms of the Cayley form is rather complicated, though certain aspects of the reconstruction are described in an appendix (and proved in, e.g. {\cite{SK}}). We need only note here that the volume form of such a $Spin(7)$ manifold can be written as
\be
{\mbox{vol}}_8 \; =\; \frac{1}{14} *\Psi \wedge \Psi \; =\; \frac{1}{14} \Psi \wedge \Psi \; ,
\label{vol}
\ee
where ${\mbox{vol}}_8 = *1$ has the single component $\sqrt{{\mbox{det}} g}$.

The closure of the form $\Psi$, eq. (\ref{dpsi}), is equivalent to the vanishing of all torsion classes of ${\bf{M}}_8$. In the forthcoming discussion we will consider only this case. However, one can relax this assumption to define more general $Spin(7)$ structures, where $d \Psi$ is expressed in terms of the non-zero torsion classes of ${\bf{M}}_8$. The associated metrics no longer have $Spin(7)$ holonomy and are not Ricci-flat but arise naturally when considering solutions of the Einstein equations in the presence of non-vanishing fluxes.

%%%%%%%%%%%%%%%%%%%%%%%%%%%%%%%%%%%%%%%%%%%%%%%%%%%%%%%%%%%%%%%%%%%%%%%%%%%%%%%%%%%%%%%%%%%%%%%%%%%%%%%%%%%%%%%%%%%%%%%%%%%%%%%%%%%%%%%%%%%%%%%%%%%%%%%%%%%

\subsection{\large{Relation to seven-manifolds with ${\mbox{G}}_2$ structure}}

In \cite{Hitchin} Hitchin showed that one can reconstruct certain special holonomy metrics as extrema of action functionals written in terms of stable forms. The stability condition is analogous to non-degeneracy of the metric for the case of general
$p$-forms. It ensures that the volume measure in the action integral is nowhere vanishing. For a general eight-manifold ${\bf{M}}_8$, there exist stable 3-form $\a$ and 5-form $\ha=*\a$ (with stabiliser $PSU(3)$). The action functional
constructed from these forms is
\be
V(\a) \; =\; {3 \over 8} \int_{{\bf{M}}_8} \ha \wedge \a \; .
\ee
For variations
\be
\a \; =\; \a_0 + d \b  \; , \quad d \a_0 \; =\; 0 \; ,
\ee
within the fixed cohomology class $[\a_0] \in H^3({\bf{M}}_8, {\mathbb{R}})$, where $\beta$ is an arbitrary 2-form, the critical points of the action above are
\bea
d \a &=& 0 \nonumber \\
d \ha &=& 0 \; .
\eea

However, the extrema of this action are {\emph{not}} $Spin(7)$ manifolds. In fact, the geometry is encoded
in the above equations in a complicated way and, in the single case solved by Hitchin \cite{Hitchins}, the manifolds  associated with these solutions are not even Ricci-flat.
In the following, we will find it more convenient to write an action explicitly in terms of the Cayley 4-form, which naturally encodes the $Spin(7)$ structure.\footnote{One might also try to localize on $Spin(7)$ geometries by a reduction and constrained variation of the stable-form action.  However, we note
that the stable 3 and 5-forms do not naturally encode $Spin(7)$ geometries.}

We now consider $Spin(7)$ eight-manifolds with the topology
\be
{\bf{M}}_8 \; =\; {\bf{M}}_7 \times {\bf {M}}_1 \; .
\ee
 If ${\bf{M}}_1$ is an interval then such eight-manifolds can be foliated by equidistant hypersurfaces (each diffeomorphic to ${\bf{M}}_7$) labeled by the coordinate $x \in {\bf{M}}_1$. Theorem 7 in \cite{Hitchin} implies that the restriction $G(x)$ of the Cayley 4-form $\Psi$ on each ${\bf{M}}_7$ hypersurface evolves as the gradient flow of the seven-dimensional action functional
\be
V_H (G) \; =\; \int_{{\bf{M}}_7} * G \wedge G \; .
\label{vol7}
\ee
The extremum of $V_H (G)$ within the cohomology class of a given $G$ determines a metric of $G_2$ holonomy on the corresponding ${\bf{M}}_7$ hypersurface. Furthermore the Cayley form can be written as
\be \label{Cayley}
\Psi \; =\; dx \wedge * G(x)+ G(x) \; ,
\label{psi}
\ee
which is closed as a result of the flow equation $d*G = \partial G / \partial x$. The converse of the construction above also follows. That is, given a closed stable 4-form $G(x)$ on ${\bf{M}}_7$ which evolves as the gradient flow of $V_H (G)$ along ${\bf{M}}_1$ (restricted to the class of $G$ in $H^4 ( {\bf{M}}_7 , {\mathbb{R}})$) then the 4-form ({\ref{psi}}) defines a metric with holonomy $Spin(7)$ on ${\bf{M}}_7 \times {\bf{M}}_1$.  In the following we take the extra dimension to be a circle, i.e.
${\bf{M}}_1 = {\bf{S}}^1$.

Unlike the $G_2$ holonomy case in seven dimensions, one already faces a difficulty in describing a $Spin(7)$ holonomy eight-manifold as the extremum of a form action because of the required self-duality condition, $\Psi = *\Psi$. Taking the usual approach that self-duality is to be imposed by hand at the level of the field equations, it seems natural to write the action functional
\be
V(\Psi) \; =\; {1 \over 2}  \int_{{\bf{M}}_8} * \Psi \wedge \Psi \; .
\label{actionp}
\ee
For generic 4-form field strength $\Psi$ however, one cannot simply obtain the equation of motion $d *\Psi = 0$ from the variational principle. The reason is that a general 4-form is not stable in eight dimensions (irrespective of whether it is self-dual or not) which means that the volume form $*\Psi \wedge \Psi$ vanishes for some values of $\Psi$. Then clearly the extrema of the action (\ref{actionp}) will be given not only by $4$-forms that satisfy the field equations but also by $4$-forms for which $*\Psi \wedge \Psi$ has zeros. To make the variational principle well-defined (i.e. localizing the extrema of the action on the field equations), one needs some way to exclude the degenerate points from the space of all 4-forms. A possibility is to define some kind of restricted variation of $\Psi$ which keeps it in a subspace of $\Lambda^4 (\bf{M}_8)$ that does not contain degenerate points. Although generically we do not know how to do this, for the special case of interest there is a natural way of implementing it. Namely, we can define the off-shell continuation of the Cayley form $\Psi$ to be a 4-form of the form given by equation (\ref{Cayley}), where the field strength $G$ is not necessarily coclosed. In particular, we only consider variations $\delta \Psi := dx \wedge \delta (* G)+ \delta G$, with $\delta G = d \Gamma$ being the variation of $G$ within a fixed cohomology class in $H^4 ( {\bf{M}}_7 , {\mathbb{R}} )$ for a given value of $x$.\footnote{A possible alternative approach to localize on $Spin(7)$ geometries may be via a constrained variation (i.e. including Lagrange multipliers) of an action containing $*\Psi \wedge \Psi$ and $\Psi \wedge \Psi$. We comment more on that at the end of Section 5.}

Substituting the explicit form of ${\Psi}$ given in ({\ref{psi}}), the action (\ref{actionp}) reduces to
$$
V(\Psi) \; =\; \int_ {\bM_7 \times {\bf S}^1 } dx \wedge *G (x) \wedge G (x) \; .
$$
The extrema of this action correspond to closed and coclosed 4-forms $G$ which determine a $G_2$ holonomy metric on ${\bf{M}}_7$, that is independent of the value of $x$.\footnote{Recall that $d*G=0$ implies $\partial G / \partial x = 0$ due to the flow equation $d*G=\partial G / \partial x$.} Our restricted variation of $\Psi$ has therefore effectively reduced the theory in eight-dimensions to the topological M-theory on seven-manifolds of $G_2$ holonomy.
Consequently the $Spin(7)$ eight-manifold ${\bf{M}}_8$ becomes a direct product ${\bf{M}}_7 \times {\bf{S}}^1$, as opposed to a non-trivial fibration.

The eighth dimension (parameterised by $x$) therefore seems to be just a spectator in the discussion above, and in particular it remains classical. Its importance will be seen in the next section where it will enable us to construct a two-torus that is transverse to the six-dimensional topological string target space in a classically topological eight-dimensional theory. The impossibility of having a full quantum theory for the Cayley form $\Psi$ may be just the counterpart in the topological context of the fact that F-theory is not on the same footing as M-theory. That is, F-theory is not a genuine quantum theory in twelve dimensions but just a technical tool that is useful for obtaining new classical solutions of type IIB superstrings.

%%%%%%%%%%%%%%%%%%%%%%%%%%%%%%%%%%%%%%%%%%%%%%%%%%%%%%%%%%%%%%%%%%%%%%%%%%%%%%%%%%%%%%%%%%%%%%%%%%%%%%%%%%%%%%%%%%%%%%%%%%%%%%%%%%%%%%%%%%%%%%%%%%%%%%%%%%%

\section{\large{Torus reduction}}

To make connection with the topological A- and B-models, we now reduce along one more circle direction (with coordinate $y$) via
the Hamiltonian flow equations considered in \cite{M}:
\bea
G &=& \hr \wedge dy + \hf k \wedge k \nonumber \\
* G &=& \rho +   k \wedge d y \; ,
\eea
where $k$ is a two-form related to the K\"{a}hler structure, and $\rho$ defines the almost complex structure of the six-manifold
$$
\Omega \; =\;  \rho + i \hr ( \rho ) \; .
$$
Hence the torus reduction of the Cayley form gives
\be
\Psi \; =\; dx \wedge  \rho - dy \wedge \hr + d x \wedge dy \wedge k + \hf k \wedge k
\label{psired} \; .
\ee
The action then effectively reduces to the sum of the symplectic and holomorphic actions
in six dimensions that were argued in \cite{M} to describe the A- and $B+\bar{B}$ models respectively:
\be \label{redac}
V(\Psi) \; =\; \int_{{\bf{M}}_6 \times {\bf{T}}^2} dx\wedge dy \wedge \left( {1 \over 2}
k \wedge k \wedge k - \hr \w \rho \right) \; .
\ee
We recall that the first term is viewed as an action for a stable four-form field strength $\sigma = \frac{1}{2} k\wedge k$. Extremizing this action with respect to variations $\delta \sigma = d \alpha$ in the fixed cohomology class $[ \sigma ] \in H^4 ({\bf{M}}_6, {\mathbb{R}})$, one obtains the field equation $dk=0$ that describes K\"{a}hler geometry. Similarly, viewing the second term as an action for a 3-form field strength $\rho$ and varying it in the fixed cohomology class $[ \rho ] \in H^3 ({\bf{M}}_6, {\mathbb{R}})$, one finds the equation of motion $d\hat{\rho}=0$, which describes complex geometry.

As a last remark in this section, we recall that the compatibility conditions for $SU(3)$ structure
\be \label{compat}
k \wedge \rho \; =\; 0 \qquad {\rm and} \qquad \frac{2}{3} \, k\wedge k\wedge k \; =\; \hat{\rho} \wedge \rho \; ,
\ee
are interpreted from the topological string perspective as a nonperturbatively generated coupling between the A- and B-model \cite{M}.

%%%%%%%%%%%%%%%%%%%%%%%%%%%%%%%%%%%%%%%%%%%%%%%%%%%%%%%%%%%%%%%%%%%%%%%%%%%%%%%%%%%%%%%%%%%%%%%%%%%%%%%%%%%%%%%%%%%%%%%%%%%%%%%%%%%%%%%%%%%%%%%%%%%%%%%%%%%

\section{{\large S-duality}}

We will now analyse the $SL(2,{\mathbb{Z}})$ modular transformations of the extra torus, which keep the Cayley form $\Psi$ and hence also the action $V( \Psi )$ invariant. Under an $SL(2,{\mathbb{Z}})$ transformation, the torus coordinates $(x,y)$ transform such that
\be
\left(\begin{array}{c} dx \\ dy \end{array}\right) \; \rightarrow\; \left(\begin{array}{cc} a & b \\ c & d  \end{array}\right)
\left(\begin{array}{c} dx \\ dy  \end{array}\right) \; ,
\ee
for any integers $a$, $b$, $c$, $d$ which obey $ad -bc =1$.

Defining
$$
d X \; :=\; \left(\begin{array}{c} dx \\ dy \end{array}\right) \; ,
\nonumber
$$
and
$$
\Xi \; :=\; \left(\begin{array}{c} {\hat \rho} \\ \rho \end{array}\right) \; ,
$$
and introducing the $SL(2,{\mathbb{Z}})$-invariant
$$
J \; =\; \left(\begin{array}{cc} 0
 & 1 \\ -1 & 0  \end{array}\right) \; ,
$$
implies the Cayley form can be written as
\be
\Psi \; =\; dX^t \wedge J \Xi + {1 \over 2}  dX^t \wedge J dX \wedge  k + {1 \over 2}
k \wedge  k \; .
\label{psired2}
\ee
The expression above is manifestly $SL(2,{\mathbb{Z}})$-invariant provided $\Xi$ transforms like $dX$, i.e. as
\be
\left(\begin{array}{c} {\hat \rho} \\ \rho \end{array}\right)
\; \rightarrow\; \left(\begin{array}{cc} a & b \\ c & d  \end{array}\right)
\left(\begin{array}{c} {\hat \rho} \\ \rho \end{array}\right)
\; ,\quad\; ad -bc =1 \; ,
\label{xitra}
\ee
in the fundamental representation of the modular group, with
\be
k \; \rightarrow\; k \; ,
\ee
transforming as a singlet.

In the canonical quantization of topological M-theory on ${\bf{M}}_6 \times {\mathbb{R}}$ considered in \cite{M}, it was found that fluctuations of the stable closed 4-form $G$ in seven dimensions (within the fixed cohomology class $[G]$) correspond to a phase space parameterized by the canonically conjugate variables $\hat\rho$ and $\rho$. Their commutation relation is
\be
\left\{ \hr , \rho \right\} \; =\;  \int_{\bM_6} \hr \wedge \rho \; .
\label{quant}
\ee

These are also the canonically conjugate variables of the B-model wavefunction. Here we find
that changing the modular parameter of the torus, while keeping the Cayley form invariant, changes the almost complex structure of ${\bf{M}}_6$, defined by
$\hat\rho$ and $\rho$, by an $SL(2,{\mathbb{Z}})$ transformation. Such $SL(2,{\mathbb{Z}})$ transformations correspond to a subgroup of the infinite-dimensional group $W_{\infty}$ of two-dimensional area preserving diffeomorphisms which leave the left and right hand side of ({\ref{quant}}) invariant.

In particular, the $SL(2,{\mathbb{Z}})$ generator S which exchanges the cycles of the torus transforms the conjugate variables $\hr
\rightarrow \rho$, $\rho \rightarrow -\hr$ so that $\Omega \rightarrow i \Omega$. This is the conjectured in \cite{M} S-duality of the A- and B-models.
The other generator T transforms $\hr \rightarrow \rho + \hr$, $\rho \rightarrow \rho$. Of course, the general $SL(2,{\mathbb{Z}})$ transformation obtained by successive applications of these generators just mixes the conjugate variables as in ({\ref{xitra}}).

The S-duality which exchanges A- and B-models was derived from the superstring S-duality in \cite{S}. Including the RR and NS gauge fields, it was argued there that the A- and B-model three-forms
\bea
{\hat \Omega_A} &=& \Omega_A + iC_R \; , \nonumber\\
{\hat \Omega_B} &=& \Omega_B + iC_{NS}
\eea
are exchanged under it as
\be
{\hat{\Omega}}_A \; \leftrightarrow\; {\hat{\Omega}}_B \; .
\ee

Following \cite{M}, we have considered the on-shell Calabi-Yau three-fold geometry which assumes the RR and NS fluxes are zero from the superstring perspective. We have also taken a unit string coupling, as in \cite{M}. We will comment more on the inclusion of coupling dependence later on. Imposing these conditions in the considerations of \cite{S} implies the relations
\bea
{\hat \Omega_A} &=& \Omega \; =\; \rho + i \hr \nonumber \\
{\hat \Omega_B} &=& i \Omega \; =\; -\hr + i \rho \; .
\eea
Hence the torus S transformations of our variables are in agreement with the prediction that S-duality exchanges the A- and B-models on the same manifold. That is, since the holomorphic 3-form of a given Calabi-Yau three-fold of fixed volume is only determined up to $U(1)$ multiplication then the S transformation above does not change this volume.

More generally, in the quantization of the B-model wavefunction or in the canonical quantization in \cite{M} one must consider fluctuating off-shell geometries. Then, in the quantization of fluctuations of $G$ within a fixed cohomology class, $\delta \hr$ is not necessarily closed and is the same type of quantity as the RR-flux. It therefore naturally couples to the worldvolume of A-model branes (i.e. Lagrangian 3-cycles). One can similarly argue a coupling for the B-model branes (i.e. holomorphic cycles). Thus S-duality has a very interesting effect in the quantum theory, where
it can generate new couplings for branes, as already predicted in \cite{M, S}.

The action of S-duality and the couplings to branes would be more transparent in geometries with background fluxes. In the following section, we discuss Hitchin's construction for form actions associated with these geometries in six dimensions. In order to lift them to eight dimensions so that we could study S-duality in a manner similar to our considerations above, we would need to generalize Hitchin's gradient flow equations to the case of non-zero flux. We leave this for future work, and here will only consider constructions of the six-dimensional form actions. Thus the forthcoming discussion is intended to be an initial step towards understanding target space actions and S-duality for these geometries.

Before proceeding with this discussion, we conclude the present section by noting a curious symmetry of the Cayley form $\Psi$ in eight dimensions
which commutes with the S-duality described above. When written as in ({\ref{psired2}}), $\Psi$ is also invariant under the transformations
\bea
\Xi &\rightarrow& \Xi + E \wedge k \nonumber \\
k &\rightarrow& k - dX^t  \wedge J E + \frac{1}{2} \, {\tilde{e}} \, dX^t \wedge J dX \; ,
\label{sym2}
\eea
provided the $SL(2,{\mathbb{Z}})$-doublet of 1-forms $E := {\, {\hat{e}} \, \choose \, e \,}$ on ${\bf{M}}_6$ and
$SL(2,{\mathbb{Z}})$-singlet 0-form ${\tilde{e}}$ on ${\bf{M}}_6$ are related such that
\be
-\frac{1}{2} \, E^t \wedge JE \; =\; {\tilde{e}} \, k \; .
\ee
This equation simply implies that ${\tilde{e}}$ is proportional to $*_6( E^t \wedge JE \wedge k \wedge k)$ so that $E$
are the only independent parameters in the symmetry transformation ({\ref{sym2}}). The particular $SL(2,{\mathbb{Z}})$ representations of the
parameters are chosen so that ({\ref{sym2}}) is compatible with the S-duality transformations described previously.

The transformation of $k$ in ({\ref{sym2}}) is slightly peculiar in that it maps a 2-form on ${\bf{M}}_6$ to a 2-form on the full space ${\bf{M}}_6 \times {\bf{T}}^2$.
Thus the transformations ({\ref{sym2}}) are not a symmetry of the reduced six-dimensional action.
Nonetheless, the symmetry in eight dimensions has an intriguing structure which mixes the A- and B-model data $k$ and $\Xi$ in a non-trivial way. It is worthwhile investigating whether this is a reflection in the topological setup of dualities of the physical superstring theories.

%%%%%%%%%%%%%%%%%%%%%%%%%%%%%%%%%%%%%%%%%%%%%%%%%%%%%%%%%%%%%%%%%%%%%%%%%%%%%%%%%%%%%%%%%%%%%%%%%%%%%%%%%%%%%%%%%%%%%%%%%%%%%%%%%%%%%%%%%%%%%%%%%%%%%%%%%%%

\section{{\large Background fluxes in six dimensions}}

In recent years it has become clear that one of the long-standing problems of string theory, namely moduli stabilization, can at least partially be resolved by compactifing on non-K\"{a}hler manifolds. The reason is that this generates a superpotential in the low energy effective theory and so some of the moduli fields get fixed. Unlike the Calabi-Yau case, which is a purely geometric compactification, the non-K\"{a}hler manifolds are solutions of superstring theory only in the presence of non-zero background fluxes.

It is well-known that in Calabi-Yau compactifications some quantities of physical interest (namely,
F-terms) can be computed using topological string theory. The worldsheet  description of the latter is in terms of a supersymmetric sigma-model with a Calabi-Yau three-fold target space and with an appropriate twisting of the worldsheet fields. Similarly, one expects that topological strings on non-K\"{a}hler manifolds may provide valuable information for the corresponding physical non-K\"{a}hler compactifications. In this regard, several topological sigma-models with non-K\"{a}hler target spaces have been considered recently \cite{kapustin}. In the present section we will write down form actions
according to Hitchin's construction in the presence of non-zero background fluxes, i.e.
actions which would be the effective action of topological strings on non-K\"{a}hler manifolds.

Let us first recall the action functional description of nearly K\"{a}hler manifolds \cite{Hitchins}. These are a particular subset of $SU(3)$ structure manifolds characterized by non-vanishing first torsion class.
Recall that the $SU(3)$ structure manifolds are classified in terms of  five torsion classes ${\cal W}_i$, $i=1,...,5$ \cite{CCDL}. Complex manifolds have ${\cal W}_1 = 0 = {\cal W}_2$. It has been shown that supersymmetry requires the internal manifold to have ${\cal W}_1 = 0$ both in type II \cite{GMPT} and heterotic \cite{Strom} compactifications\footnote{More precisely, for IIB and heterotic superstrings, the internal space has to be complex whereas for IIA superstrings it must be twisted symplectic.}. So nearly K\"{a}hler manifolds do not seem to be of immediate physical interest.

 On the other hand, in certain cases, in the large complex structure limit, one can consider superstrings on half-flat manifolds\footnote{These are $SU(3)$ structure manifolds whose intrinsic torsion belongs to ${\cal W}_1^- \oplus {\cal W}_2^- \oplus {\cal W}_3$, where $"{}^-"$ denotes the imaginary part of the corresponding class. Equivalently, they can be defined by requiring $k\wedge dk = 0$ and $d\Omega^- = 0$ while $dk \neq 0$, $d\Omega \neq 0$ (see \cite{CCDL}).} to be a good approximate description of the low energy effective theory \cite{GLMW}. The nearly K\"{a}hler manifolds are a subset of the half-flat ones and in \cite{Micu} they were even argued to capture important information about the resulting superpotential. So it is conceivable that topological strings on them are of some interest too. With this motivation in mind, let us review the constrained variational problem whose critical points give these manifolds \cite{Hitchins}. Consider the action functionals
\be \label{nK}
V_1 (\rho, \sigma) = \int_{{\bf{M}}_6} \Big( \rho \wedge \hat{\rho} + \frac{1}{2} k\wedge k\wedge k\Big) \, , \qquad V_2(\rho,\sigma) = \int_{{\bf{M}}_6} d\alpha \wedge \beta \, ,
\ee
where $\rho = d\alpha \in \Omega^3 ( {\bf{M}}_6 )$ and $\sigma = \frac{1}{2} k\wedge k = d\beta \in \Omega^4 ( {\bf{M}}_6 )$. The field equations obtained by varying $\alpha$ and $\beta$, while keeping $V_2 = 1$, are
\be \label{Feq}
d \hat{\rho} = - \lambda \,k\wedge k \, , \qquad d k = \lambda \,\,\rho \, ,
\ee
where $\lambda$ is a Lagrange multiplier. The constrained variation is necessary to enforce non-degeneracy of the functional $\int_{{\bf{M}}_6} d\alpha \wedge \beta$. The action $V_1$ in (\ref{nK}) is just the sum of the actions for the holomorphic 3-form and K\"{a}hler form encountered in section 3. Equations (\ref{Feq}) define an $SU(3)$ structure manifold with ${\cal W}_1 \neq 0$ and ${\cal W}_{2,3,4,5} = 0$, i.e. a nearly K\"{a}hler manifold. The role of the first torsion class is played by the Lagrange multiplier $\lambda$. Putting $\lambda = 0$, one recovers the Calabi-Yau case.

We note that the compatibility conditions for $SU(3)$ structure (\ref{compat}) follow from (\ref{Feq}), up to rescalings. Indeed, taking the derivative of the first equation in (\ref{Feq}) and using the second one results in
\be \label{rwk}
\rho \wedge k =0 \, ,
\ee
for $\lambda \neq 0$. Using the above relation and again (\ref{Feq}) one also finds the second condition in (\ref{compat}) \cite{Hitchins}. So we see here that these compatibility conditions, that previously had to be imposed on the unification of the topological A- and B- models as constraints arising non-perturbatively\footnote{In \cite{M} they were shown to follow from the lift to a seven-dimensional $G_2$ manifold.}, can be automatically incorporated in the six-dimensional action, at least for some non-K\"{a}hler compactifications.

It is very interesting to understand how the above  construction can be generalized to non-K\"{a}hler manifolds with ${\cal W}_{1,2}=0$ and ${\cal W}_{3,4,5} \neq 0$, which are of much greater physical relevance\footnote{For a very non-exhaustive list of references on the vast subject of flux compactifications, see \cite{GMW}.}. Here we make initial steps in that direction. A well-defined action functional which localizes on manifolds with ${\cal W}_{4,5} \neq 0$ is
\be \label{S1}
V(\rho, \sigma) = \int_{{\bf{M}}_6} \left[ \hat{\rho}\wedge \rho - \frac{1}{2} k\wedge k\wedge k + \mu \left( \hat{\rho}\wedge \rho -\frac{2}{3} k\wedge k\wedge k \right) \right]\, ,
\ee
where $\mu$ is a smooth function. By varing $\rho$ and $\sigma$ in a fixed cohomology class, i.e. by taking
\be
\rho = \rho_0 + d\alpha \, , \qquad \sigma = \sigma_0 + d\beta
\ee
for some fixed closed 3- and 4-forms $\rho_0$, $\sigma_0$, and arbitrary 2- and 3-forms $\alpha$, $\beta$, we find the field equations
\be
d\hat{\rho} = - d \ln(\mu+1)\wedge \hat{\rho} \, , \qquad dk = - d \ln\left(\frac{4}{3}\mu+1\right) \wedge k \, .
\ee
Thus the action above is indeed related to non-K\"{a}hler manifolds with
\be
{\cal W}_4 = - d \ln\left(\frac{4}{3}\mu+1\right)\, , \qquad {\cal W}_5 =- d \ln(\mu+1) \, .
\ee
Note that the $\mu \rightarrow 0$ limit gives the Calabi-Yau action with zero torsion classes. A nice feature of (\ref{S1}) is that the field equation for $\mu$ enforces the relation between the volumes determined by $\rho$ and $k$ that is the second compatibility condition for $SU(3)$ structure manifolds in  (\ref{compat}).

Another action that seems very natural to consider in order
to incorporate the physically interesting NS and RR fluxes is
\be \label{flac}
V(\rho, \sigma) = \int_{{\bf{M}}_6} \left( (\rho+C)\wedge(\hat{\rho}+{\hat{C}}) + (k+B)\wedge (k+B)\wedge (k+B) \right) \, ,
\ee
where we can identify $\hat{C}$ with the RR 3-form potential $C_R$ that couples to the D2-branes of the topological A model, whereas $C$ $-$ with the potential\footnote{The existence of this 'NS'-type potential was predicted from mirror symmetry in \cite{Vafa}.} that couples to the B model NS2-branes of \cite{NV}. The two-form $B$ can be taken to be a combination of the NS B-field, $B_{NS}$, which
couples naturally to the fundamental string F1; and the 2-form potential $B_{R}$, under which the holomorphic D1-branes of the B model are charged. Under S-duality \cite{S}, these branes are exchanged as follows: D2 $\leftrightarrow$ NS$2$, D1 $\leftrightarrow$ F1.

Invariance of the first term in (\ref{flac}) under $SL(2,{\mathbb{Z}})$ transformations can be achieved by requiring that the doublet $(C, \hat{C})$ transform the same way as $(\rho, \hat{\rho})$. The second term is more subtle as $k$ is a singlet under our $SL(2,{\mathbb{Z}})$. Taking $B = B_{NS} + B_R$ would ensure that this term is invariant under the S transformation since S-duality is expected to exchange $B_{NS} \leftrightarrow B_R$. The issue of the full $SL(2,{\mathbb{Z}})$ invariance is related to the role of the coupling constant which is yet to be understood.

Now, varing $\rho$ in a fixed cohomology class we obtain
\be
d\hat{\rho} = -d\hat{C} \, ,
\ee
whereas varing $\sigma=\frac{1}{2} k\wedge k$:
\be
\frac{1}{2} dk = - dB - d*(B \wedge B) \, .
\ee
So the action (\ref{flac}) is related to non-K\"{a}hler manifolds whose nonvanshing torsion classes are determined by the decomposition of $d(B+*(B \wedge B))$ and $d\hat{C}$ into $SU(3)$ representations.
Although the stability properties of this action have to be investigated more thoroughly, each term in it seems very natural. The first one, $(\rho+C)\wedge (\hat{\rho}+\hat{C})$, is inspired by the combinations $\hat{\Omega}_A = \Omega_A + i C_R$ and $\hat{\Omega}_B = \Omega_B + i C_{NS}$ introduced in \cite{S}, whereas the second by Hitchin's construction of generalized Kahler manifolds from an action functional \cite{NH}.

The transformation of  $(C, \hat{C})$ is also very natural from the point
of view of our eight-dimensional torus S-duality, where one expects a simple change $\rho \rightarrow
\rho + C$ and ${\hat \rho} \rightarrow \hat{\rho} + \hat{C}$.
To make this precise, we should construct the proper eight dimensional lift i.e. a generalization of Hitchin's flow
including background fluxes.
 Clearly, the form actions for geometries with background fluxes and their possible lifts to seven and eight dimensions deserve further study and we hope to come back to this topic in the near future.

As a last remark, we note that Hitchin's constrained variations provide a general procedure of making well-defined actions, which might have had degeneracies otherwise. This suggests that there might be a way of defining an 8-dimensional action, whose equations of motion, obtained from arbitrary fixed cohomology class variations of $\Psi$, would determine a $Spin(7)$ structure metric, by considering terms like $\Psi\wedge \Psi$ and $\Psi\wedge * \Psi$ and introducing suitable Lagrange multipliers. This possibility may be worth further investigation.

%%%%%%%%%%%%%%%%%%%%%%%%%%%%%%%%%%%%%%%%%%%%%%%%%%%%%%%%%%%%%%%%%%%%%%%%%%%%%%%%%%%%%%%%%%%%%%%%%%%%%%%%%%%%%%%%%%%%%%%%%%%%%%%%%%%%%%%%%%%%%%%%%%%%%%%%%%%

\section{\large{Summary and discussion}}

In this note we found evidence for the existence of a topological theory on $Spin(7)$ eight-manifolds, which
would be analogous to physical F-theory. The topological F-theory considered was constructed on the product manifold ${\bf{M}}_6 \times {\bf{T}}^2$. We found that $SL(2,\mathbb{Z})$ transformations of the torus,
which keep the Cayley form invariant, induce $SL(2,\mathbb{Z})$ transformations of the real
and imaginary part of the 3-form $\Omega$ that were the canonically conjugate variables in the naive quantization in \cite{M}.

 An important point which remains to be resolved is
the embedding of this duality in the full superstring theory. In particular, it was discussed in
\cite{S} how deriving the topological S-duality from the S-duality of type IIB superstrings naturally leads to the inclusion
of the string coupling constant. As it is derived from the full superstring theory, topological S-duality
should invert the string coupling constant
 $g \rightarrow 1/g$. In particular, since the A-model and B-model are related by S-duality
 on the same Calabi-Yau manifold, we have the relations $g_A = 1 /g_B$ and $g_{\mu \nu}^A =  g_{\mu \nu}^B /g_B$.
 Transforming only the holomorphic 3-form $\Omega$, but not $\bar{\Omega}$, this implies that
 $k_A = k_B/g_B$ and $\rho_A = \rho_B /g_B^3$. In our formulation, since the holomorphic and
 antiholomorphic part of the volume form are treated on equal footing, such a scaling appears unnatural
 to introduce (although it is certainly natural from the perspective of the B-model Kodaira-Spencer theory).

For a better understanding of the action of topological S-duality, it seems essential
 to include background fluxes in the geometry. As a starting point for these investigations we have considered form actions, based on Hitchin's formalism, which localize on six-dimensional generalized geometries with fluxes. The higher dimensional lift for these actions and
 the study of S-duality on these generalized backgrounds are both interesting directions for further investigation.

In particular, a starting point for understanding the inclusion of the coupling in our considerations can be the construction of a six-dimensional action with background fluxes that is invariant under the S transformation. For example, the
 IIB S-duality should be used to write an invariant combination of $B_R$, $B_{NS}$ and the coupling $g$.
 In addition, powers of the coupling should also be introduced in the terms containing the potentials $(C, {\hat C})$ along the lines of \cite{S}.
 We leave these delicate and important issues for future work
\footnote{It is worth noting though that one can obtain the correct $1/g$ coupling \cite{S} in the six-dimensional action from our framework
by simply reducing on a torus with metric $|dx + i/g \, dy|^2$ (rather than the square torus metric $|dx+idy|^2$ we have considered).
Of course, these two metrics are related by diffeomorphism and replacing $y$ with $y/g$ in ({\ref{psired}}) and ({\ref{redac}})
indeed reproduces the desired coupling dependence.
In physical F-theory this choice of metric would just correspond to having a non-vanishing dilaton whilst keeping the axion set to zero.  }
.

 One may also hope to relate the coupling constant $g$ to the size of the extra circle as in
 the relation between physical M-theory and IIA superstrings, along the lines suggested in \cite{M}. The difficult point here
 is that by reducing the extra circle one actually obtains both A and $B+{\bar{B}}$ theories,
 and so it is not clear which coupling should be related to the size of the circle,
 and whether there is a physical basis for such a relation.
The inclusion of couplings as described above and the lift of the full action to eight dimensions
seem to be essential for such a physical interpretation.

  Finally, it would also be interesting to consider open topological string Chern-Simons like
  target space actions in seven and eight dimensions, which could then form the basis for a D-brane
  interpretation. Investigations along these lines have recently appeared in \cite{Z}.

%%%%%%%%%%%%%%%%%%%%%%%%%%%%%%%%%%%%%%%%%%%%%%%%%%%%%%%%%%%%%%%%%%%%%%%%%%%%%%%%%%%%%%%%%%%%%%%%%%%%%%%%%%%%%%%%%%%%%%%%%%%%%%%%%%%%%%%%%%%%%%%%%%%%%%%%%%%

\subsection*{Acknowledgment}

The work of L.A. and P.dM. is supported in part by DOE grant DE-FG02-95ER40899.
A.S. would like to thank the Michigan Center of Theoretical Physics for an invitation which
initiated this work. The work of A.S. is supported by the Stichting FOM.

%%%%%%%%%%%%%%%%%%%%%%%%%%%%%%%%%%%%%%%%%%%%%%%%%%%%%%%%%%%%%%%%%%%%%%%%%%%%%%%%%%%%%%%%%%%%%%%%%%%%%%%%%%%%%%%%%%%%%%%%%%%%%%%%%%%%%%%%%%%%%%%%%%%%%%%%%%%

\section*{\large{Appendix : metric reconstruction from the Cayley form}}

The explicit reconstruction of the $Spin(7)$ holonomy metric from the Cayley 4-form is rather complicated. For seven-dimensional $G_2$ manifolds one simply proceeds by contracting the indices of the Levi-Civita symbol with seven of the 9 indices of the tensor product of three invariant 3-forms so as to obtain a second rank symmetric tensor proportional to the $G_2$ metric. It is straightforward to see algebraically that one cannot make a metric in the same way in eight dimensions via contractions of Levi-Civita symbols with Cayley forms.

The purpose of this appendix is to write the norm defined by the reconstructed $Spin(7)$ metric (following the analysis in {\cite{SK}}). Given a vector $v$ on a $Spin(7)$ eight-manifold with Cayley form $\Psi$, it is convenient to define
\bea
A(v) &:=& ( \iota_v \Psi )_{abc} \Psi_{defg} \, \varepsilon^{abcdefg} \nn
B_{ij} (v) &:=& ( \iota_v \Psi )_{iab} ( \iota_v \Psi )_{jcd} ( \iota_v \Psi )_{efg} \, \varepsilon^{abcdefg} \; ,
\eea
where $\iota_v$ denotes the eight-dimensional interior product with $v$ and $\varepsilon$ is the $SL(7,{\mathbb{R}})$-invariant Levi-Civita symbol. Then Theorem 4.3.5 in {\cite{SK}} states that the norm is given by
\be
|v|^2 \; =\; v^t g v \; =\; c \left( \frac{( \det_7 (B_{ij} (v)) )^{1/6}}{(A(v))^{3/2}} \right) \; ,
\ee
for any 8-vector $v$ (the value of the constant $c$ is given in {\cite{SK}}).

%%%%%%%%%%%%%%%%%%%%%%%%%%%%%%%%%%%%%%%%%%%%%%%%%%%%%%%%%%%%%%%%%%%%%%%%%%%%%%%%%%%%%%%%%%%%%%%%%%%%%%%%%%%%%%%%%%%%%%%%%%%%%%%%%%%%%%%%%%%%%%%%%%%%%%%%%%

%
\end{document}